\documentclass{article}
\usepackage{latexsym}
\usepackage{graphicx}
\usepackage{epstopdf}
\usepackage{amsmath}
\usepackage{amssymb}
\begin{document}
\begin{center}
\LARGE
\textbf{Entanglement Swapping in the Transactional
Interpretation of Quantum Mechanics}\\[1cm]
\large
\textbf{Louis Marchildon}\\[0.5cm]
\normalsize
D\'{e}partement de chimie, biochimie et physique,\\
Universit\'{e} du Qu\'{e}bec,
Trois-Rivi\`{e}res, Qc.\ Canada G9A 5H7\\[0.2cm]
(louis.marchildon$\hspace{0.3em}a\hspace{-0.8em}\bigcirc$uqtr.ca)\\
\end{center}
\medskip
%
%
%
%
\begin{abstract}
The transactional interpretation of
quantum mechanics, which uses retarded
and advanced solutions of the Schr\"{o}dinger
equation and its complex conjugate, offers an
original way to visualize and understand quantum
processes.  After a brief review,
we show how it can be applied to different
quantum situations, emphasizing the importance
of specifying a complete configuration of
absorbers.  We consider in more detail
the phenomenon of entanglement swapping, and
see how the apparent retroactive enforcement of
entanglement can be understood in the
transactional interpretation. 
\end{abstract}
\section{Introduction}
It was more than 25 years ago that
John Cramer proposed what he called
the transactional interpretation (TI) of
quantum mechanics, in a comprehensive
paper~\cite{cramer} that is still an
excellent introduction to the subject.
In Cramer's view, ``interpretation must
not only relate the formalism to physical
observables.  It must also define the
domain of applicability of the formalism
and must interpret the nonobservables
in such a way as to avoid paradoxes and
contradictions.''  Indeed his approach
was not meant to be empirically different
from standard quantum mechanics, but rather
to make it more understandable.

The transactional interpretation in fact
proposes a mechanism to visualize quantum
processes.  In Cramer's words, it
``provides a description of the state vector
as an actual wave physically present in real
space and provides a mechanism for the
occurrence of nonlocal correlation effects
through the use of advanced waves.''

The idea of interpreting the wave function
as a real wave goes back to
Schr\"{o}dinger~\cite{schrodinger}
and de Broglie~\cite{broglie}.  It is
further developed in Khrennikov's more
recent ``Prequantum classical statistical
field theory'' \cite{khrennikov1,khrennikov2},
where references to other related approaches
can be found.

In spite of its intuitive appeal, TI has not
hitherto raised the amount of interest of
better known interpretations like de
Broglie-Bohm mechanics~\cite{broglie,bohm} or
Everett's relative states~\cite{everett,saunders},
let alone the Copenhagen
interpretation~\cite{heisenberg,jammer}.
For instance the year 2013 appears to be the
first time when it has been discussed at one of
the V\"{a}xj\"{o} conferences.  Hopefully the
recent monograph~\cite{kastner} published on the
subject will contribute in raising awareness of~TI.

Cramer's original proposal has since undergone
several developments, such as involving a
hierarchy of transactions~\cite{cramer2} or the
reality of possibility~\cite{kastner}.  Other
contributions have stayed closer to the original
approach and the block-universe picture of
time~\cite{marchildon1,boisvert}.

In section~2 we briefly review TI and its
``explanation'' of the Born rule.  Section~3
analyzes several quantum situations that have been
proposed as challenges to~TI.  The importance of
fully specifying the configuration of absorbers is
emphasized.  In section~4 we examine the
phenomenon of entanglement swapping which, to
our knowledge, has not hitherto been
investigated from the point of view of TI.  We show
how the apparent retroactive enforcement of
entanglement can be understood in~TI.  We
conclude in section~5.
%
\section{The transactional interpretation}
Cramer's transactional interpretation of quantum
mechanics was inspired by the
time-symmetric electromagnetic theory
of Wheeler and Feynman~\cite{wheeler1,wheeler2}.
In this approach advanced solutions of Maxwell's
equations are as physically significant
as retarded ones, and the
universe is a perfect absorber of all the
electromagnetic radiation emitted in it.
In the spirit of Wheeler and Feynman,
Cramer attributes physical reality both
to solutions of the Schr\"{o}dinger equation
(propagating forward in time) and to their complex
conjugates (propagating backward in time).

The relevance of the complex conjugate
solutions comes from the fact that the
Schr\"{o}dinger equation should be viewed as
the nonrelativistic limit of a relativistically
invariant equation (like Dirac's or
Klein-Gordon's), where advanced solutions
occur naturally.  In this sense Cramer's
theory is relativistically invariant.
A field-theoretical generalization of it
is outlined in~\cite{kastner}.

An example of a quantum-mechanical process is
the emission of a microscopic particle (e.g.\
an electron or a photon)
at some time $t_0$, followed by its absorption at a
later time $t_1$. The usual solution $\psi$
of the Schr\"{o}dinger equation (called by Cramer
an ``offer wave''), which originates at $t_0$,
propagates through $t > t_0$. This solution
$\psi$ reaches all potential detectors.
Its amplitude $\psi ( \mathbf{r}_i, t_i)$ at
detector $i$ is given by the Schr\"{o}dinger equation.
Each detector in turn emits an advanced (or
``confirmation'') wave, propagating backwards in time.
Cramer argues that the confirmation
wave coming from detector $i$ reaches the
source with an amplitude proportional to
\begin{equation}
|\psi ( \mathbf{r}_i , t_i )|^2
= \psi ( \mathbf{r}_i , t_i ) \psi^* ( \mathbf{r}_i , t_i ) .
\label{born}
\end{equation}
The first factor on the right-hand side
of~(\ref{born}) coincides
with the amplitude of the stimulating offer wave at $i$,
while the second one comes from the fact that the
confirmation wave develops as the time reverse of the
offer wave. Note that all confirmation waves
coming from all possible detectors reach the source
at the same time $t_0$. Under the constraint
that all relevant conservation laws are
satisfied, a ``transaction'' is eventually
established between the emitter and one
of the detectors, which corresponds to a 
completed quantum-mechanical process.
If the probability that the transaction is
established with detector $i$ is taken to be
proportional to the amplitude~(\ref{born})
of the confirmation wave coming from that detector,
Born's rule follows.
%
\section{Applications and challenges}
\subsection{Renninger's experiment}
\begin{figure}[h]
\centering
\includegraphics[width=5cm]{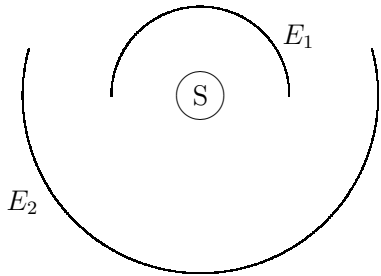}
\caption{Renninger's experiment}
\label{fig1}
\end{figure}

Renninger's negative-result 
experiment~\cite{renninger}, an example
analyzed by Cramer~\cite{cramer}, provides a
nice illustration of how a transaction works.
Figure~\ref{fig1} depicts a stationary
source~S at the center of two truncated
spherical shells $E_1$ and $E_2$ of radii
$R_1$ and $R_2$, respectively.  Shell $E_1$
subtends a solid angle of $2\pi$ whereas
shell $E_2$ subtends at least the
complementary solid angle.  On each shell,
perfect absorbers are lined up that will
detect any particle coming from~S.
The system is set up so that at time $t_0$
the source emits exactly one particle with
speed approximately equal to $v$ and
spherically symmetric wave function.
Let $t_1 = t_0 + R_1/v$.  For $t_0 < t < t_1$,
the state vector can be represented~as
\begin{equation}
|\psi\rangle = \frac{1}{\sqrt{2}}
\left(|E_1\rangle + |E_2\rangle \right) ,
\label{renn}
\end{equation}
where each vector $|E_i\rangle$
is associated with detection at
the corresponding shell $E_i$.

Suppose that at some time $t > t_1$
no detector at $E_1$ has fired.
In standard quantum mechanics this means that
the particle's state vector has collapsed
at $t_1$ and is thereafter equal to $|E_2\rangle$.
But in Cramer's interpretation there
is no such thing as a collapse occurring
at a specific time $t_1$ or later at
$t_2 = t_0 + R_2/v$.  Instead there is
completion of a transaction. The whole process
is viewed atemporally in four-dimensional
space-time (Cramer says in ``pseudotime''),
along the full interval
between emission and absorption. The condition that
only one particle is emitted at the source translates into
the fact that only one transaction is established in the
end, either between~S and $E_1$ (with probability
$1/2$) or between~S and $E_2$.
%
\subsection{Maudlin's challenge}
Advanced interactions may raise the specter
of inconsistent causal loops.  This
doesn't happen in Cramer's theory, owing
to the lack of independent control
on advanced waves.  But Maudlin has
argued~\cite{maudlin} that they can lead to
inconsistent evaluations of probabilities.

\begin{figure}[h]
\centering
\includegraphics[width=7cm]{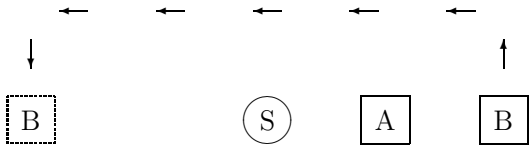}
\caption{Maudlin's challenge}
\label{fig2}
\end{figure}

Maudlin's challenge is depicted in
figure~\ref{fig2}.  A slow particle is emitted
by source~S to the left or to the right, with
equal probability.  Two detectors are
initially stationed on the right-hand
side, one behind the other.  If the particle
is not detected by~A, then~B quickly swings
to the left in time to catch it.  Maudlin
observes that every time B goes to the left,
it absorbs the particle.  Yet the amplitude
of the confirmation wave arriving at the
source from the left-hand side in this case
is equal to~0.5 only.  How can absorption
by~B always occur in that situation?

An answer to Maudlin's challenge has been
proposed in~\cite{marchildon1}, and
assumes as in Wheeler-Feynman
electrodynamics that every offer wave is
eventually absorbed.  This is equivalent to
putting a third detector (say~C) at the
far left.  Then the probability of
absorption by any detector to the left
of~S becomes equal to the amplitude of the
confirmation wave coming from the left.
The scheme is fully consistent with the
block-universe picture of time.

Several other answers to Maudlin's challenge
have also been proposed and are thoughtfully
reviewed in~\cite{lewis}.  Berkovitz~\cite{berkovitz}
points out that in situations involving causal
loops, relative frequencies of events may not
coincide with objective chances.  Cramer~\cite{cramer2}
suggests ``a \emph{hierarchy of transaction
formation}, in which transactions across small
space-time intervals must form or fail before
transactions from larger intervals can enter
the competition.''  Kastner~\cite{kastner}
introduces the framework of a ``possibilist
transactional interpretation'' (PTI), where
the offer and confirmation waves are viewed as
real, but not actual, waves of possibility.
And Lewis~\cite{lewis} argues for considering
emitters and detectors as full-fledged quantum
systems, a view strongly criticized by
Kastner~\cite{kastner1}.
%
\subsection{Interaction-free measurements}
\begin{figure}[h]
\centering
\includegraphics[width=7cm]{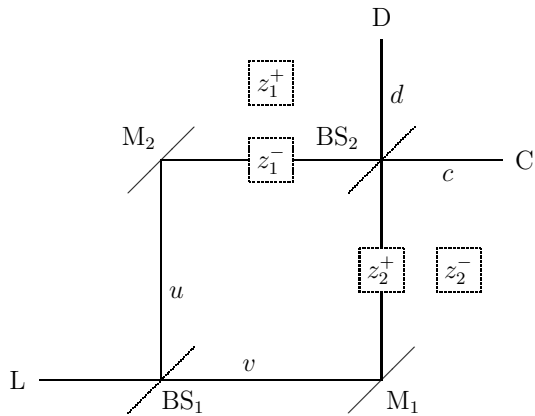}
\caption{The quantum liar experiment}
\label{fig3}
\end{figure}

Another challenge which has been raised
against Cramer's ideas comes from the
so-called quantum liar
experiment~\cite{elitzur}, shown
in figure~\ref{fig3}.  Photons are emitted
by the laser~L, and directed to a
Mach-Zehnder interferometer.  In each
path of the interferometer is an atom
which can absorb a photon if it interacts
with it.  The atoms are in a superposition
of spin plus and minus states with respect
to the $z$~direction.

In each run of the experiment, the
photon-atoms state vector can be followed
through beam splitter~BS$_1$, reflection
at~M$_1$ and~M$_2$, entanglement of the
photon with either atom~1 or atom~2,
recombination at BS$_2$ and, finally,
possible absorption by~C or~D
\cite{boisvert,elitzur}.  It follows directly
from the expression of the
final state vector that if a photon
is absorbed by detector~D, then the atoms are
left in an entangled state of the form
\begin{equation}
|\psi\rangle_{\text{atoms}} = \frac{1}{\sqrt{2}}
\left( |z_1^+ \rangle |z_2^+ \rangle
+ |z_1^- \rangle |z_2^- \rangle \right) . 
\end{equation}
The problem consists in
explaining how one photon, necessarily
absorbed if it interacts with an atom,
can entangle both of them.

Kastner~\cite{kastner,kastner2} approaches
this problem in the framework of PTI, where
offer and confirmation waves ``have access to
a larger physically real space of possibilities.''
This is motivated by the observation that
$N$-particle wave functions are defined in
$3N$-dimensional configuration space which,
from a physical point of view, has no
straightforward projection into ordinary
3-dimensional space.

On the other hand, a careful analysis of
absorbers~\cite{boisvert} again
provides a solution closer to the original
formulation of TI.  Eventual measurements
of the atoms' spin necessary to verify
consequences of their entanglement require
detectors that are specifically set to measure
the spin projection on the $z$~axis.  Moreover,
absorption at~D entails that the confirmation
wave coming from~D will go through both paths.
Atoms~1 and~2 are linked through the interplay
of offer waves interacting with them
\emph{en route} to~D and confirmation waves going
from~D to them, which explains the
entanglement.
%
\section{Entanglement swapping}
\subsection{The problem}
The process of entanglement
swapping~\cite{zukowski} consists in
entangling two quantum systems that have
never interacted, through each system's
entanglement with another one.  It can
be illustrated as follows.

Suppose that Alice can prepare
entangled spin~$1/2$ particles labelled~1
and~2.  It will be useful to define
``Bell states'' $|\Phi^{\pm}\rangle_{12}$
and $|\Psi^{\pm}\rangle_{12}$ as
\begin{align}
|\Phi^{\pm}\rangle_{12} &= \frac{1}{\sqrt{2}}
\left( |+\rangle_1 |+\rangle_2
\pm |-\rangle_1 |-\rangle_2 \right) , \\
|\Psi^{\pm}\rangle_{12} &= \frac{1}{\sqrt{2}}
\left( |+\rangle_1 |-\rangle_2
\pm |-\rangle_1 |+\rangle_2 \right) ,
\end{align}
where tensor product signs are understood
between adjacent kets and all spin components
are defined with respect to the same axis.
We also assume that Bob, far away from Alice,
can prepare similar states labelled~3 and~4.
For simplicity, the spatial parts of the
state vectors are suppressed throughout.
A more complete treatment would also
associate offer and confirmation waves
with them.  See~\cite{kastner}, Appendix~C,
for a situation where such considerations
are relevant.

At some time $t_0$, Alice prepares
her particles in the singlet state 
$|\Psi^{-}\rangle_{12}$ and Bob prepares
his in the state $|\Psi^{-}\rangle_{34}$.
The four-photon initial state
is therefore given by
\begin{align}
|\Psi\rangle_{1234} 
&= |\Psi^{-}\rangle_{12} |\Psi^{-}\rangle_{34} \notag\\
&= \frac{1}{2} \left(
|+\rangle_1 |-\rangle_2 |+\rangle_3 |-\rangle_4
- |+\rangle_1 |-\rangle_2 |-\rangle_3 |+\rangle_4
\right. \notag\\ & \qquad \left.
- |-\rangle_1 |+\rangle_2 |+\rangle_3 |-\rangle_4
+ |-\rangle_1 |+\rangle_2 |-\rangle_3 |+\rangle_4
\right) .
\label{ent1}\end{align}
It is simple algebra to show that this
can also be written as
\begin{align}
|\Psi\rangle_{1234} &= \frac{1}{2} \left(
|\Psi^{+}\rangle_{14} |\Psi^{+}\rangle_{23}
- |\Psi^{-}\rangle_{14} |\Psi^{-}\rangle_{23}
\right. \notag\\ & \qquad \left.
- |\Phi^{+}\rangle_{14} |\Phi^{+}\rangle_{23}
+ |\Phi^{-}\rangle_{14} |\Phi^{-}\rangle_{23}
\right) .
\label{ent2}\end{align}

Now suppose that Alice sends particle~2
and Bob sends particle~3 to Eve.  Then
Eve can make spin measurements in her
Bell-state basis $|\Psi^{\pm}\rangle_{23}$ 
and $|\Phi^{\pm}\rangle_{23}$.
If she gets the result $|\Psi^{+}\rangle_{23}$,
one sees from~(\ref{ent2}) that particles~1
and~4 become entangled in state
$|\Psi^{+}\rangle_{14}$ without ever having
interacted.

It was pointed out in~\cite{peres} that
the measurement of particles~2 and~3 can
be made after measurements have been carried
out on particles~1 and~4.  If the measurements
of~2 and~3 are made in the Bell state basis,
then~1 and~4 behave as in entangled states.
If, however, the measurements
of~2 and~3 are made in the basis
\begin{equation}
\left\{ |+\rangle_2 |+\rangle_3,\;
|+\rangle_2 |-\rangle_3,\; 
|-\rangle_2 |+\rangle_3, \;
|-\rangle_2 |-\rangle_3 \right\} ,
\label{basis}\end{equation}
then~1 and~4 behave as in product states.
In the words of~\cite{ma},
``whether [the] two photons are entangled
(showing quantum correlations)
or separable (showing classical correlations)
can be defined after they have been measured.''

Different interpretations of quantum
mechanics can address this paradox in
various ways.  We will show in the next
section how TI characteristically helps to
visualize the situation.
%
\subsection{The view from TI}
\begin{figure}[h]
\centering
\includegraphics[width=7cm]{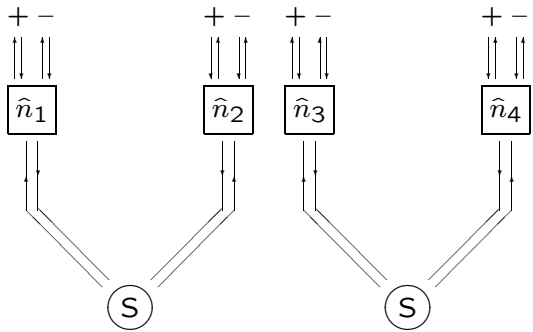}
\caption{Offer and confirmation waves
for product state measurement}
\label{fig4}
\end{figure}

In TI the pattern of offer and confirmation
waves depends on the configuration of
detectors.  Let us first consider
the case where particles~2 and~3 are
measured in basis~(\ref{basis}).
The situation is depicted in
figure~\ref{fig4}. Here offer waves are
represented by upward arrows,
confirmation waves by downward arrows
and all measurement axes coincide
($\hat{n}_1 = \hat{n}_2 = \hat{n}_3 = \hat{n}_4$).

It is seen that
the arrows of particle~2 do not connect
with the ones of particle~3.  Hence there
can be no connection between the arrows
of~1 and~4, and thus no entanglement.

\begin{figure}[h]
\centering
\includegraphics[width=7cm]{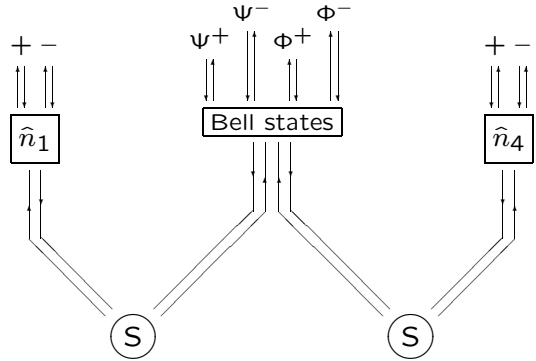}
\caption{Offer and confirmation waves
for Bell state measurements}
\label{fig5}
\end{figure}

In the situation where measurements of~2
and~3 are made in the Bell state basis,
the pattern of offer and confirmation waves
is depicted in figure~\ref{fig5}.

Since Bell states are linear combinations
of states of particles~2 and~3, the
corresponding offer and confirmation waves
(represented by upward and downward arrows,
respectively) are superpositions of offer
and confirmation waves of both~2 and~3.
Whichever Bell state results from a quantum
measurement, this establishes an unbroken
link between particles~2 and~3 and, through
the sources, between particles~1 and~4.
This connection is responsible for the
entanglement between~1 and~4.  More explicitly,
a confirmation wave originating from a detector
set to measure particle~1 will reach the
source on the left.  It connects to an offer
wave leaving the source towards the right,
which reaches the Bell state detectors.
Because the Bell states are superpositions,
their detectors emit confirmation waves towards
both sources, in particular the one to the right.
The offer wave from that source reaches detectors
associated with particle~4.  The upshot is that
in TI, the presence or absence of entanglement
between~1 and~4 depends on the full
configuration of offer and confirmation
waves, itself dependent on the specific
configuration of detectors.  Whether there is
entanglement or not is defined
early, but it is causally dependent on the
future configuration of detectors.
%
\section{Conclusion}
The transactional interpretation of quantum
mechanics provides a way to visualize quantum
phenomena.  The fact that it involves advanced
causation has led some to believe that it
gives rise to paradoxes.  But the full
specification of offer and confirmation waves,
crucial in TI, necessitates a complete
specification of detectors.  This is
consistent with the block-world view of
space-time and provides a general framework
for the solution of quantum paradoxes.
\section*{Acknowledgements}
I am grateful to the Natural Sciences and
Engineering Research Council of Canada for financial
support.  I thank Stevie Turkington for 
bringing the interest of entanglement swapping
to my attention.
%
%

%
\end{document}